\documentclass{moriond}

\def\Journal#1#2#3#4{{#1} {\bf #2}, #3 (#4)}

\def\PRD{{\em Phys. Rev.} D}

\def\be{\begin{equation}}
\def\ee{\end{equation}}
\def\bea{\begin{eqnarray}}
\def\eea{\end{eqnarray}}

\newcommand{\Photo}{}

\begin{document}
\vspace*{2cm}
\title{An influence of the matter distribution on the positional accuracy of reference sources}

\author{ T.I. Larchenkova (1), A.A.Lutovinov (2), N.S.Lyskova (3,2)}

\address{(1) ASC of P.N.Lebedev Physical Institute, Leninskiy prospect 53, Moscow, Russia\\
(2) Space Research Institute, Profsoyuznaya 84/32, Moscow, Russia\\
(3) MPA, Karl-Schwarzschild-Str. 1, 85748 Garching, Germany
}

\maketitle\abstracts{
We consider an influence of a non-stationary gravitational field of the Galaxy on the visible
positions of extragalactic sources. A contribution of the baryonic component of the galactic matter as well
as of the hidden matter (including a population of brown dwarfs) were took into account.
The observed variations of the deflection angle of light rays in a gravitational field of randomly
moving point-like masses can be considered as a stochastic process. Using such an approach we
constructed an autocorrelation function of studied stochastic process and found that its relative
changes are about 15\% for one year and about 35\% for ten years.
}

In the nearest future technologies will allow one to carry out extremely accurate radio
interferometrical observations with an angular resolution of 1\,$\mu$as and optical observations with the
accuracy of 10\,$\mu$as per year. Such an accuracy requires to take
into account effects of the general relativity, arising during the propagation of electromagnetic waves in
non-stationary gravitational fields. In this respect it is very important to estimate an amplitude of changes of
observed positions (coordinates) of extragalactic sources due to the propagation of their emission in the
non-stationary gravitational field both visible stars in the galactic disk/bulge and invisible massive
objects in the halo.

Measured source coordinates are the random functions of time, therefore variations of the light-ray deflection 
angle from the direction between the source and observer can be considered as a stochastic process. This process
can be described by several statistical values -- the mathematical expectation, dispersion and correlation
(or autocorrelation) function of the process under consideration.

The electromagnetic signal from a distant extragalactic source is propagated in the gravitational field of $N$
arbitrary moving point-like bodies with different masses. To simplify our calculations we assume that
velocities of deflecting bodies do not change in time and the smallest distance between the photon and a
deflecting body is much smaller compared to any other distances in the system. Then
the deflection angle of the light by the $a$-th body from the straight line equals \cite{KS}:

\begin{equation}
\alpha_{a}^i(t)=\frac{4Gm_{a}}{c^2}\frac{1-\vec {k}\vec {v_{a}}}{\sqrt {1-\vec {v_{a}}^2}}\frac{P_{j}^i r_{a}^j}{|P_{j}^i r_{a}^j|^2},
\end{equation}
where $\vec k$ is the unit vector, directed from the source to the observer,
$\vec x_{a}, m_{a}, \vec v_{a}$ -- coordinates, mass, velocity of the $a$-th deflecting body, respectively,
$\vec r_{a}$ -- the distance between the point $\vec x(t)$ at the photon trajectory and the $a$-th deflecting body,
$c$ -- the speed of light in vacuum, $G$ -- the gravitational constant, $P_{ij}=\delta_{ij}-k_{i}k_{j}$ --
the operator of the vector projection on the plane, which is perpendicular to $\vec k$.

\begin{figure}
\centerline{\includegraphics[width=0.45\linewidth,bb=47 185 566 672]{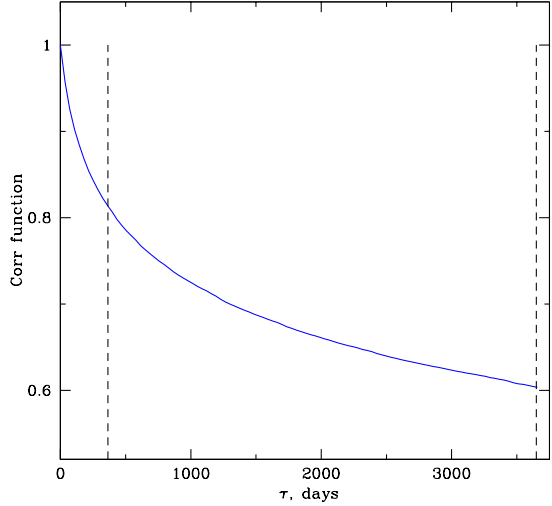}}
\caption[]{The autocorrelation function. Dashed lines mark one and ten years periods.}
\label{fig:autocorr}
\end{figure}

A deflection of electromagnetic signals in the non-stationary gravitational field of the Galaxy is a random
process $\varepsilon (t)$:

\begin{equation}
\varepsilon (t)=\alpha(t) - \langle \alpha(t) \rangle,
\end{equation}
where $\langle \alpha(t) \rangle=0 $. For a stationary noise process, the autocorrelation function depends not on the individual
times $t_i$ and $t_j$, but only on their difference $\tau$ and can be written as

\begin{equation}
\Re (t,\tau)= \int dm_a d\vec x_a d\vec v_a f(m_a, \vec x_a, \vec v_a) \varepsilon (t,m_a,\vec x_a, \vec v_a)
\varepsilon (t+\tau,m_a,\vec x_a, \vec v_a),
\end{equation}

\noindent where we assume that the statistical ensemble of deflection bodies is defined by uncorrelated parameters,
so the distribution function can be approximated by the product of three statistically independent distribution functions~\cite{LK2006}

\begin{equation}
f(m_a, \vec x_a, \vec v_a) \propto f(m_a) f(\vec x_a) f(\vec v_a).
\end{equation}

We use an exponential function of the Kroupa model \cite{Kroupa}
as the mass distribution function of deflection bodies. Let us consider two model density distributions:
a multi-component model of the Galaxy \cite{DB} which include population of brown dwarfs \cite{Chabrier}
and a four-component model of the Galaxy \cite{BS,Bahcall}.
The velocity distribution is a truncated Maxwellian profile.

The calculated autocorrelation function is shown in Fig.\ref{fig:autocorr}, its changes are about 15\% for
one year and about 35\% for ten years.

\section*{Acknowledgments}

TL acknowledge the support by the grant NSh-4235.2014.2, AL acknowledge the support by the grant RFBR 13-02-12094.

\end{document}